\documentclass[%
 aip,
 amsmath,amssymb,
 reprint,%
]{revtex4-1}

\usepackage{graphicx}
\usepackage{dcolumn}
\usepackage{bm}

\usepackage[utf8]{inputenc}
\usepackage[T1]{fontenc}
\usepackage{mathptmx}
\usepackage{etoolbox}
\usepackage{chemformula} 
\makeatletter
\def\@email#1#2{%
 \endgroup
 \patchcmd{\titleblock@produce}
  {\frontmatter@RRAPformat}
  {\frontmatter@RRAPformat{\produce@RRAP{*#1\href{mailto:#2}{#2}}}\frontmatter@RRAPformat}
  {}{}
}%
\makeatother
\begin{document}

\preprint{AIP/123-QED}

\title[]{Spin-orbital entanglement in Cr$^{3+}$-doped glasses}

\author{J. S. Robles-P\'aez}
\affiliation{Escuela de Qu\'imica, Universidad Industrial de Santander, Cra. 27--9, 680002 Bucaramanga, Colombia}
\author{A. T. Carre\~no-Santos}
\affiliation{Escuela de Qu\'imica, Universidad Industrial de Santander, Cra. 27--9, 680002 Bucaramanga, Colombia}
\author{V. Garc\'ia-Rojas}
\affiliation{Escuela de Qu\'imica, Universidad Industrial de Santander, Cra. 27--9, 680002 Bucaramanga, Colombia}
\author{J. F. P\'erez-Torres}
\email{jfperezt@uis.edu.co}
\affiliation{Escuela de Qu\'imica, Universidad Industrial de Santander, Cra. 27--9, 680002 Bucaramanga, Colombia}


\begin{abstract}
	A framework for reconstructing the one-electron spinors, $\Gamma_7$ and $\Gamma_8$, of \ch{Cr^3+} ions embedded
	in glasses from optical measurements has been developed. These spinors provide the basis for calculating the
	spin-orbital von Neumann entropy, offering a quantitative measure of quantum entanglement within the electronic
	state. To illustrate the applicability of this concept, an aluminum phosphate glass doped with 1 mol$\%$ chromium
	was prepared and characterized via optical absorption spectroscopy. By extracting the fundamental electronic
	parameters, including the spin-orbit coupling constant $\xi_{\rm 3d}$, the crystal field strength $Dq$, and
	the Racah parameters $B$ and $C$, we demonstrate how the spin-orbital entanglement entropy, $\Delta S_{\rm vN}^{\rm SO}$,
	can be mapped across different chemical environments. Our analysis reveals that while individual crystal field
	parameters do not dictate the degree of entanglement, the dimensionless ratio between the spin-orbit coupling
	and the crystal field strength ($\xi_{\rm 3d}/Dq$) exhibits a robust linear correlation with the entropy.
	This relationship serves as a clear illustration of how the competition between relativistic effects and
	local symmetry governs the information content of the 3d($O_h$) electronic manifold.
\end{abstract}

\maketitle

\section{Introduction}
\begin{figure}[h!]
	\centering
	\includegraphics[width=0.25\textwidth]{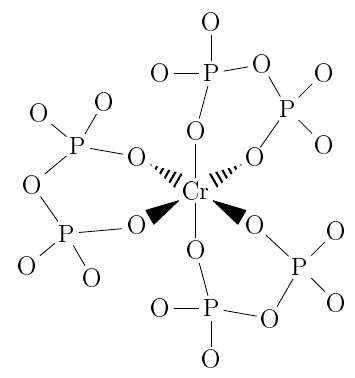}
	\caption{Idealized octahedral structure of CrO$_6$ in phosphate glasses.}\label{fig:CrPO}
\end{figure}

Quantum entanglement was originally conceived as a phenomenon where two or more particles become
so deeply linked that their quantum states cannot be described independently, regardless of the
distance separating them. This phenomenon has been observed in several scenarios using photons
\cite{Freedman1972,Aspect1982}, electrons \cite{Hensen2015}, helium atoms \cite{Athreya2026},
and calcium monofluoride molecules \cite{Holland2023}. The concept extends beyond multi-particle systems
to include different degrees of freedom within a single entity. Specifically, it is now possible
to address the entanglement between the spin and orbital angular momentum of a single electron within
an atom, molecule, or solid \cite{Takayama2021,Gotfryd2024,Minarro2024,Shen2026}.
In transition metal ions, particularly those with an odd number of electrons, this entanglement drives
the magnetic properties of 4d and 5d systems \cite{Takayama2021}, where spin-orbit coupling (SOC) is significant.
Nevertheless, certain 3d systems can also exhibit appreciable spin-orbital entanglement in specific environments.
The importance of spin-orbital interaction in transition metal-halide perovskites has been recently identified
\cite{Anandan2023}. Among the compounds with 3d transition metal ions that display spin-orbital
entanglement are glasses doped with \ch{Cr^3+}. In such systems, spin-orbital interaction manifests
itself in the visible absorption spectrum, where one or two absorption bands show an interference
pattern. Moreover, some phosphors such as \ch{RbAl3P6O20} doped with \ch{Cr^3+} can exhibit the
interference pattern  in the photoluminescence spectrum \cite{Wu2025}, and materials such as \ch{Mg2SnO4}
doped with \ch{Cr^3+} exhibit the interference pattern in the thermoluminescence spectrum \cite{Xie2021}.
The optical absorption spectrum of transition metal ions embedded in a glass can be explained in terms
of the crystal field theory. For example, in a phosphate glass doped with \ch{Cr^3+}, the chromium(III)
is surrounded by six oxygen ions in octahedral symmetry, see Figure \ref{fig:CrPO}. In this case, the
crystal field theory predicts the $\rm 3d$ atomic orbitals split into $t_{2g}$ and $e_g$ crystal field
orbitals. The three electrons occupy the six lowest energy spin-orbitals $t_{2g}\otimes(\alpha\cup\beta)$,
that is electron configuration $t_{2g}^3$, giving rise to multiplets $\rm ^4A_2$, $\rm ^2E$, $\rm ^2T_1$
and $\rm ^2T_2$. According to the first Hund's rule, the multiplet $\rm ^4A_2$ is the ground state.
When an electron jumps to one of the four high energy spin-orbitals $e_g\otimes(\alpha\cup\beta)$, it
yields the electron configuration $t_{2g}^2e_g^1$ and the multiplets $\rm ^2A_1$, $\rm ^2A_2$, $\rm ^2E$,
$\rm ^4T_1$, and $\rm ^4T_2$ are formed. Because the spin selection rule, only the transitions
$\rm ^4T_2\leftarrow {^4A_2}$ and $\rm ^4T_1\leftarrow {^4A_2}$ are expected to be observed, accounting
for the absorption bands. However, as stated before, the absorption bands of glasses doped with \ch{Cr^3+}
show interference patterns. Bussi\`ere et al. \cite{Bussiere2003} and Maalej et al. \cite{Maalej2016}
have managed not only to explain the interference pattern in terms of the interaction between the doublets
$\rm ^2E$, $\rm ^2T_1$, and $\rm ^2T_2$ with the quadruplets $\rm ^4T_2$ and $\rm ^4T_1$, but also to
extract the one-electron spin-orbit coupling constant from the pattern based on a method developed by
Neuhauser et el. \cite{Neuhauser2000}. In this work, we employ a framework based on the relativistic
crystal field theory \cite{jperez2024} and Neuhauser's method \cite{Neuhauser2000} to quantify the
spin-orbital entanglement in chromium(III)-doped glasses. The standard measurement of entanglement is
the entanglement entropy \cite{SanzVicario2017,Blavier2022,Greene-Diniz2025,Mosquera2026}. We prove that
for glasses doped with \ch{Cr^3+}, and in general for Cr$^{3+}$ systems exhibiting interference patterns
in the absorption spectrum, the spin-orbital entanglement entropy can be calculated from the optical
absorption spectrum. The spin-orbit coupling interaction and the corresponding interference pattern have
been analyzed in Cr$^{3+}$-doped fluoride \cite{Maalej2016} and tellurite \cite{Taktak2021} glasses.
To increase the variety of glasses, we prepare a chromium(III)-doped aluminum phosphate glass and calculate
the spin-orbital entanglement from the interference pattern appearing in the absorption spectrum.
The remainder of this paper is organized as follows: In Section \ref{Theory} the crystal field theory and
its relativistic version is briefly described together with the Neuhauser et al. \cite{Neuhauser2000} and
Bussi\`ere et al. \cite{Bussiere2003} method to extract the one-electron spin-orbit coupling constant.
In Section \ref{Preparation} the preparation of the chromium-doped aluminum phosphate glass is described.
In Section \ref{Results} the results are presented, comprising the calculation and the analysis of the
spin-orbital entanglement entropy. Finally, some conclusions are outlined in Section \ref{Conclusion}.

\section{\label{Theory}Theoretical framework}
\subsection{Crystal field theory}
\begin{figure}[h]
\centering
\includegraphics[width=0.35\textwidth]{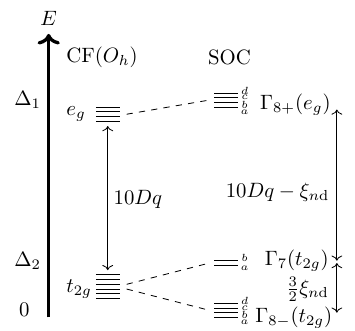}
	\caption{Energy splitting of d orbitals due to octahedral crystal field CF($O_h$) and to the spin-orbit coupling SOC.}\label{figLevels}
\end{figure}

Crystal field theory (CFT), established primarly by Bethe \cite{Bethe1929} and van Vleck \cite{vanVleck1978},
describes the optical and magnetic properties of $\rm d^n$  transition metal atoms surrounded by $N$ point
charges, or ligands, in a specific geometric configuration. In its standard formulation, perturbation theory
is used to determine the eigenvalues and eigenvectors of the crystal field Hamiltonian ${\cal H}_{\rm CF}$:
\begin{equation}
	{\cal H}_{\rm CF} = \alpha\hbar c \sum_{i=1}^{N}\frac{Z}{|{\bf r}-{\bf R}_i|}
\end{equation}
defined within the basis of atomic d-orbitals $|n{\rm d}_m\rangle$. Here, $\alpha$ is the fine structure constant,
$\hbar$ is the reduced Planck constant, $c$ is the velocity of the light in vacuum, $Z$ is the effective charge
of the ligands, ${\bf r}$ is the position of the electron, and ${\bf R}_i$ is the position of each ligand.
For six-coordinated ions, the octahedral crystal field $(O_h)$ yields the following eigenvectors \cite{BallhausenBook}:
\begin{eqnarray}
	&& |t_{2g}^-\rangle = |n{\rm d}_{-1}\rangle \\
	&& |t_{2g}^+\rangle = |n{\rm d}_{+1}\rangle \\
	&& |t_{2g}^0\rangle = \frac{1}{\sqrt{2}}\left(|n{\rm d}_{-2}\rangle - |n{\rm d}_{+2}\rangle\right) \\
	&& |e_g^a\rangle = \frac{1}{\sqrt{2}}\left(|n{\rm d}_{-2}\rangle + |n{\rm d}_{+2}\rangle\right) \\
	&& |e_g^b\rangle = |n{\rm d}_0\rangle
\end{eqnarray}
with corresponding eigenvalues: $E(e_g) = 6Dq$ and $E(t_{2g}) = -4Dq$, where $Dq$ is the crystal field strength
defined by
\begin{equation}
	Dq = \alpha\hbar c\frac{Z}{6a}\int_0^\infty |f_{n{\rm d}}(r)|^2 r^4 {\rm d}r
\end{equation}
where $a$ stands for the ligand-metal distance. The function $f_{n{\rm d}}(r)$ is the radial part of the d-atomic orbitals
$\psi_{n{\rm d}m}(r)= f_{n{\rm d}}(r)Y_{2m}(\theta\phi)/r$ with $Y_{2m}(\theta\phi)$ the spherical harmonics of $\ell=2$.
In the relativistic version of the crystal field theory \cite{jperez2024,jperez2025}, the eigenvectors of the total Hamiltonian,
incorporating the Dirac term,
${\cal H} = {c\hat{\boldsymbol \alpha}\cdot\hat{\boldsymbol p} + \beta m_{\rm e}c^2 + V(r)} + {\cal H}_{\rm CF}$ are obtained
in the basis of atomic Dirac d-spinors $|n\kappa m_j\rangle$,
\begin{eqnarray}
	|\Gamma_{8\pm}^a\rangle &=& x_\pm\left(\sqrt{\tfrac{5}{6}}|n{-3}{-\tfrac{5}{2}}\rangle
                                        +\sqrt{\tfrac{1}{6}}|n{-3}{+\tfrac{3}{2}}\rangle\right) \nonumber \\ 
				       && +y_\pm|n{+2}{+\tfrac{3}{2}}\rangle \\
	|\Gamma_{8\pm}^b\rangle &=& x_\pm\left(\sqrt{\tfrac{5}{6}}|n{-3}{+\tfrac{5}{2}}\rangle
                                        +\sqrt{\tfrac{1}{6}}|n{-3}{-\tfrac{3}{2}}\rangle\right) \nonumber \\
					&& -y_\pm|n{+2}{-\tfrac{3}{2}}\rangle \\
	|\Gamma_{8\pm}^c\rangle &=& x_\pm|n{-3}{-\tfrac{1}{2}}\rangle + y_\pm|n{+2}{-\tfrac{1}{2}}\rangle \\
	|\Gamma_{8\pm}^d\rangle &=& x_\pm|n{-3}{+\tfrac{1}{2}}\rangle - y_\pm|n{+2}{+\tfrac{1}{2}}\rangle \\
	|\Gamma_7^a\rangle &=& \sqrt{\tfrac{1}{6}}|n{-3}{-\tfrac{5}{2}}\rangle -\sqrt{\tfrac{5}{6}}|n{-3}{+\tfrac{3}{2}}\rangle \label{eq:G7a} \\
	|\Gamma_7^b\rangle &=& \sqrt{\tfrac{1}{6}}|n{-3}{+\tfrac{5}{2}}\rangle - \sqrt{\tfrac{5}{6}}|n{-3}{-\tfrac{3}{2}}\rangle \label{eq:G7b} 
\end{eqnarray}
with
\begin{eqnarray}
        && x_\pm = \frac{\delta\pm\sqrt{\delta^2+1}}{\sqrt{\left(\delta\pm\sqrt{\delta^2+1}\right)^2+1}} \\
        && y_\pm = \frac{1}{\sqrt{\left(\delta\pm\sqrt{\delta^2+1}\right)^2+1}} \\
        && \delta = \frac{1}{2\sqrt{6}(p/q)}\left(\frac{5\xi_{n{\rm d}}}{4Dq}+1\right) \label{eq:delta} \\
        && \xi_{n{\rm d}} = \frac{2}{5}\left(E(n{\rm d}_{5/2}) - E(n{\rm d}_{3/2})\right) \\
        && Dq = \alpha\hbar c \frac{Z}{6a} \int_0^\infty |F_{n{\rm d}_{5/2}}(r)|^2 r^4 {\rm d}r \\
        && p/q = \frac{\int_0^\infty F_{n{\rm d}_{5/2}}(r)F_{n{\rm d}_{3/2}}(r)r^4{\rm d}r}{\int_0^\infty |F_{n{\rm d}_{5/2}}(r)|^2 r^4 {\rm d}r}
        \label{eq:pq}
\end{eqnarray}
and eigenvalues
\begin{eqnarray}
	E(\Gamma_{8\pm}) &=& Dq\left(1 \pm \sqrt{\left(\frac{5\xi_{n{\rm d}}}{4Dq}+1\right)^2 + 24(p/q)^2}\right) \nonumber \\
			  && - \frac{\xi_{n{\rm d}}}{4} \label{eq:E8} \\
	E(\Gamma_7) &=& \xi_{n{\rm d}} - 4Dq \label{eq:E7} 
\end{eqnarray}
Here, the Dirac quantum numbers $\kappa=-3$ and $\kappa=2$ correspond to the $j=5/2$ and $j=3/2$ manifolds, and $\xi_{n{\rm d}}$ stands
for the spin-orbit coupling constant.
Unlike standard crystal field orbitals ($t_{2g}$ and $e_g$), the relativistic spinors $\Gamma_i$ depend not only on the
spin-orbit coupling constant $\xi_{n{\rm d}}$, but also on the crystal field parameter $Dq$ and on the relativistic ratio $p/q$.
This relativistic approach is essential when SOC is experimentally observable and becomes fundamental for describing 5d
transition metal ions, where spin-orbit interactions cannot be neglected \cite{jperez2024,jperez2025}. In the non-relativistic
limit, where the radial functions $F_{n{\rm d}_{3/2}}(r)$ and $F_{n{\rm d}_{5/2}}(r)$ are identical, the ratio $p/q$ reduces to
unity and $\xi_{n{\rm d}}$ to zero, thereby recovering the standard CFT results, i.e. $E(\Gamma_{8-})=E(\Gamma_7)=E(t_{2g})$,
and $E(\Gamma_{8+})=E(e_g)$, see Fig. \ref{figLevels}.
We have recently developed a methodology for quantifying the entanglement between the spin and orbital degrees of freedom in
single-electron systems \cite{GarciaRojas2025}. We remind the reader that a state is entangled if it can not be written as a
separable state, i.e. $|\Gamma_i\rangle\ne|n\ell m\rangle\otimes|sm_s\rangle$.
While \ch{Cr^3+} is a $\rm 3d^3$ multi-electron system, its low-lying electronic structure can be effectively treated within
a independent-particle framework. In this approach, we assign a single-particle spinor $|\Gamma_i\rangle$ to the each active
electron, where the complex electron-electron correlations are implicitly accounted for by the Racah parameters $B$ and $C$.
Because $B$ and $C$ (interelectronic repulsion) and $Dq$ (crystal field strength) are intrinsically coupled through the
nephelauxetic effect and the spatial expansion of the d-orbitals, they define a consistent `mean field' environment.
Consequently, the spin-orbital entanglement calculated for an individual spinor serves as a representative metric for
the correlation dynamics of the $\rm d^3$ manifold, as these parameters dictate the mixing of the independent-particle
Dirac spinors $|n\kappa m_j\rangle$ for all three electrons simultaneously. For a \ch{Cr^3+} ion in a octahedral environment
$(O_h)$, the single-particle spinor is represented by one of the symmetry-adapted spinors $\Gamma_7$ or $\Gamma_8$.
To evaluate the degree of entanglement, we utilize the von Neumann entropy, $S_{\rm vN}$. The total density matrix for the
system is defined as $\hat{\rho}=|\psi\rangle\langle\psi|$. By performing a partial trace over the orbital degreees of freedom,
we obtain the reduced spin density matrix:
\begin{equation}
	\hat{\rho}^{\rm spin}_{\Gamma_i} = {\rm Tr}_{\rm orbital}(\hat{\rho})
\end{equation}
The spin-orbital entanglement entropy is then calculated as:
\begin{equation}
	S_{\rm vN}(\Gamma_i) = -{\rm Tr}(\hat{\rho}^{\rm spin}_{\Gamma_i}\log_2(\hat{\rho}^{\rm spin}_{\Gamma_i})) 
				= -\sum_k\lambda_k(\Gamma_i)\log_2(\lambda_k(\Gamma_i)) 
\end{equation}
where $\lambda_k(\Gamma_i)$ are the eigenvalues of the reduced desnity matrix $\hat{\rho}^{\rm spin}_{\Gamma_i}$.
Since $\sum_k\lambda_k(\Gamma_i)=1$,
i.e. $\langle\alpha|\hat{\rho}^{\rm spin}_{\Gamma_i}|\alpha\rangle + \langle\beta|\hat{\rho}^{\rm spin}_{\Gamma_i}|\beta\rangle=1$
with $|\alpha\rangle$ and $|\beta\rangle$ the electron spin states, the von Neumann entropy can be obtained as:
\begin{equation}
	S_{\rm vN}(\Gamma_i) = -\lambda(\Gamma_i)\log_2(\lambda(\Gamma_i)) - (1-\lambda(\Gamma_i))\log_2(1-\lambda(\Gamma_i))
\end{equation}
with the eigenvalues of the ten spinors equal to
\begin{eqnarray}
	&& \lambda(\Gamma_7^{ab}) = \frac{2}{3} \\
	&& \lambda(\Gamma_{8\pm}^{ab}) = \frac{2}{15}x_\pm^2 + \frac{1}{5}y_\pm^2 - \frac{2\sqrt{6}}{15}x_\pm y_\pm \\
	&& \lambda(\Gamma_{8\pm}^{cd}) = \frac{2}{5}x_\pm^2 + \frac{3}{5}y_\pm^2 - \frac{2\sqrt{6}}{5}x_\pm y_\pm 
\end{eqnarray}
Due to the symmetry restrictions on the wave function imposed by the double point group $O_h^*$ \cite{CeulemansBook},
the spin-orbital entanglement entropy has to be redefined as
\begin{equation}
	\Delta S_{\rm vN}^{\rm SO}(\Gamma_i) = S_{\rm vN}(\Gamma_i) - S_{\rm vN}^{(\xi_{n{\rm d}}=0)}(\Gamma_i)
\end{equation}
otherwise spinors $\Gamma_i$ arising from $\rm 3d$, $\rm 4d$ and $\rm 5d$ systems would exhibit the same entropy.
However, only spinors $\Gamma_8^{ab}(t_{2g})$ and $\Gamma_7^{ab}(t_{2g})$ are affected by the term
$S_{\rm vN}^{(\xi_{n{\rm d}}=0)}$ \cite{GarciaRojas2025}.
This quantity provides a measure of the non-separability of the spin and orbital wavefunctions. In the context of
$\rm 3d^3$ ions like \ch{Cr^3+}, this metric allows us to quantify how the relativistic spin-orbit coupling and the
local crystal field environment cooperate to mix these degrees of freedom. This effect is not present in the
orbitals of the standard crystal field theory.
For $\rm 3d$ ions, the relativistic ratio can be approximated to one, $p/q=1$. Therefore, only the crystal field strength
$Dq$ and the spin-orbit coupling constant $\xi_{n{\rm d}}$ are needed to calculate the corresponding entropies.

\subsection{Determination of the crystal field parameters}
This section details the extraction of the Racah parameters ($B$, $C$), the crystal field strength ($Dq$), and the
spin-orbit coupling constant ($\xi_{\rm 3d}$) from the optical absorption spectrum. While the calculation of spin-orbital
entanglement entropies depends explicitly on $Dq$ and $\xi_{\rm 3d}$, we also determine the Racah parameters $B$ and $C$.
Since $Dq$, $B$, and $C$ are fundamentally interrelated, including the latter allows us to examine whether changes in
the interelectronic repulsion correlate with the degree of entanglement.
For a $\rm 3d^3$ ion in octahedral field, the crystal field strength $Dq$ and the Racah parameter $B$ can be
straightforwardly obtained by \cite{Dou1990}
\begin{eqnarray}
	&& Dq = \frac{E({\rm ^4T_2})}{10} \label{eq:Dq} \\
	&& B = \frac{(E({\rm ^4T_1})-2E({\rm ^4T_2}))(E({\rm ^4T_1})-E({\rm ^4T_2}))}{15E({\rm ^4T_1})-27E({\rm ^4T_2})} \label{eq:B}
\end{eqnarray}
while the second Racah parameter $C$ is obtained by varying it in the interaction matrix \cite{SuganoBook}
\begin{widetext}
\begin{equation}
	\begin{pmatrix}
		-12Dq-6B+3C & -6\sqrt{2}B    & -3\sqrt{2}B & 0 \\
		-6\sqrt{2}B & -2Dq+8B+7C     & 10B         & \sqrt{3}(2B+C) \\
		-3\sqrt{2}B & 10B            & -2Dq-B+3C   & 2\sqrt{3}B \\
		0           & \sqrt{3}(2B+C) & 2\sqrt{3}B  & 18Dq-8B+4C
	\end{pmatrix}\label{eq:Matrix}
\end{equation}
\end{widetext}
until the lowest eigenvalue coincides with the observed energy $E({\rm ^2E})$. In order to have a guest value for $C$,
we use the Auxiliary Tanabe-Sugano diagram\cite{Garcia2023}, suitable to estimate the crystal field strength and both Racah
parameters. When the $\rm ^2E$ and $\rm ^2T_1$ states interact with the $\rm ^4T_1$ state through the spin-orbit coupling,
the optical band $\rm {^4T_1}\leftarrow{^4A_2}$ exhibits one or two interference dips. This interference is frequently observed
in $\rm d^3$ chromium and $\rm 3d^8$ nickel ions \cite{Lempicki1980,Bussiere2003,Villain2010,Maalej2016},
and can be used to estimate an effective spin-orbit coupling constant.
Following the framework developed by Neuhauser et al. \cite{Neuhauser2000} and Bussi\`ere et al. \cite{Bussiere2003},
the one-electron spin-orbit coupling constant \cite{Maalej2016}, $\xi_{\rm 3d}=\sqrt{5/6}\gamma_1$, is obtained by
fitting the parameters $\gamma_1$, $\gamma_2$, $\Gamma_0$, $\Gamma_1$ and $\Gamma_2$ of the line shape function $\sigma(\omega)$
\begin{eqnarray}
	&& \sigma(\omega) = -\frac{1}{\pi}{\rm Im}\left(\frac{\beta}{1-(\alpha_1\gamma_1^2+\alpha_2\gamma_2^2)\beta}\right) \\
	&& \alpha_i = \frac{1}{\omega - \epsilon_i + {\rm i}\Gamma_i}, \quad i=1,2 \\
	&& \beta = \frac{1}{\omega - 10Dq + {\rm i}\Gamma_0}
\end{eqnarray}
to the exprimental optical absorption spectrum. This is achieved by minimizing the integral
\begin{equation}
	I(\Gamma_1,\Gamma_2,\gamma_1,\gamma_2) = \int_{\omega_1}^{\omega_2}
	\left[\sigma(\omega,\Gamma_1,\Gamma_2,\gamma_1,\gamma_2)-\sigma_{\rm exp}(\omega)\right]^2{\rm d}\omega
\end{equation}
Here, $\omega$ is the photon energy, $\epsilon_1$ and $\gamma_1$ are the energy at the dip and the coupling constant between states
$\rm ^2E$ and $\rm ^4T_2$, respectively; $\epsilon_2$ and $\gamma_2$ are the energy at the dip and the coupling constant between states
$\rm ^4T_2$ and $\rm ^2T_1$, respectively; $\Gamma_0$, $\Gamma_1$ and $\Gamma_2$ are damping factors that determines the width of
individual vibronic lines of the spin allowed $\Gamma_0$ and spin forbidden $\Gamma_1$ and $\Gamma_2$ transitions, respectively.
For more datails on the parameters $\gamma_i$ and $\Gamma_i$ reader can consult Refs. \cite{Bussiere2003,Maalej2016}.

\section{\label{Preparation}Preparation of the glass system}
Conventional melt-quenching technique was used to prepare aluminum phosphate glass with 50P$_2$O$_5$-44Na$_2$O-5Al$_2$O$_3$-Cr
composition using as raw materials reagent grade Al$_2$O$_3$ (neutral, chromatography grade, Carlo Elba), Na$_2$CO$_3$ ($\ge99.9~\%$,
Sigma-Aldrich) and NH$_4$H$_2$PO$_4$ ($\ge98~\%$, Sigma-Aldrich).
The choice of tris(acetylacetonato)chromium(III), \ch{Cr(acac)3}, as a precursor allows for a chloride-free synthesis,
preventing unintended alterations in the glass composition. Beyond its chemical stability, which ensures the exclusive
incorporation of Cr$^{3+}$, this complex serves as a critical benchmark; recent studies using time-resolved L-edge X-ray
spectroscopy have characterized its metal-centered excited states, specifically the $\rm{^2E}\leftarrow{^4A}$ spin-flip transition
\cite{Ghodrati2025}. Cr(acac)$_3$ was synthesized according to the procedure reported by Fernelius et al. \cite{Fernelius1957}
using CrCl$_3\cdot6$H$_2$O ($\ge99~\%$, J. T. Baker) and acetylacetone ($\ge99~\%$, Sigma-Aldrich) on an Anton Paar monowave
400 microwave synthesis reactor. Regarding the glass preparation, appropriate quantities of the compounds were weighed, mixed,
and ground in an agate mortar. The mixture was then melted in porcelain crucibles under air atmosphere using an electric furnace
programmed to increase the temperature at $\rm 10~^\circ C/min$ from room temperature to $\rm 950~^\circ C$ and maintained at
this temperature for 1 hour.
Rate of $\rm 10~^\circ C/min$ was used in order to eliminate water, ammonia and acac$^-$. The melt was poured rapidly onto
a preheated graphite mold at $\rm 350~^\circ C$ and then cooled down to room temperature.
The fabricated glass piece, measuring $1.9\times1.0\times0.5~{\rm cm}$, was weighed and found to be 2.033~{\rm g}.
The absorption spectrum was recorded using a Genesys 50 UV-Vis spectrometer (Thermo Fisher Scientific).

\section{\label{Results}Results}
\begin{figure*}[t]
\centering
\includegraphics[width=0.85\textwidth]{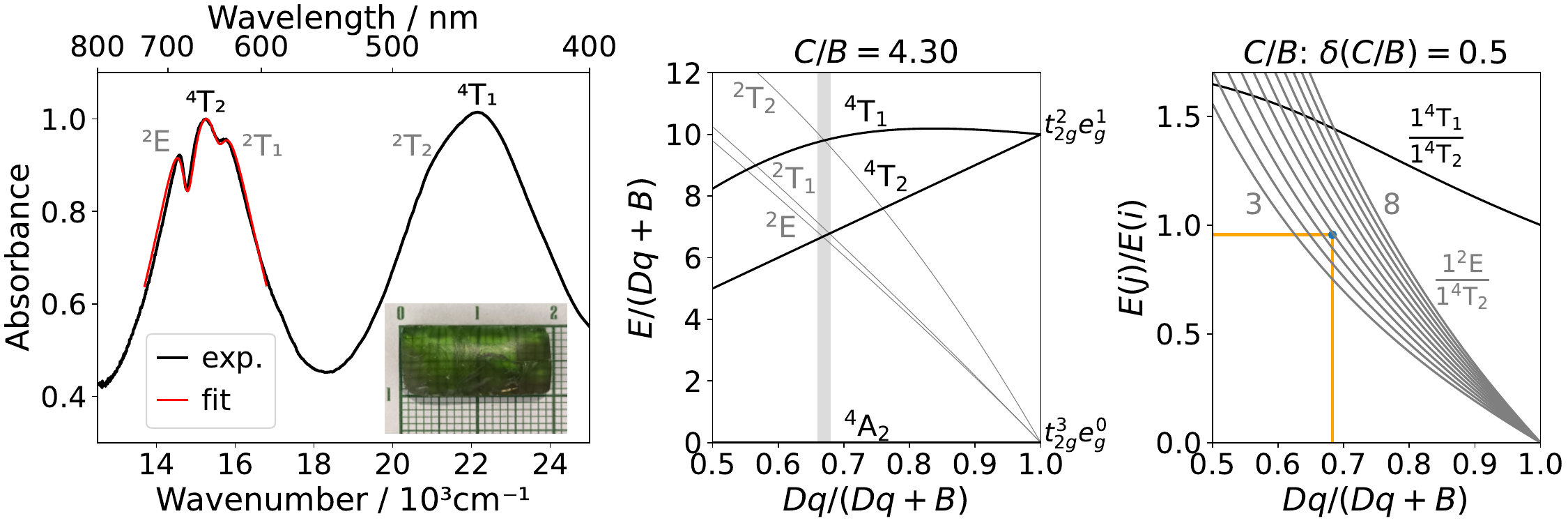}
	\caption{Left: absorption spectrum of $\rm 1~mol~\%$ $\rm Cr^{3+}$ doped phosphate glass. Inset: photo of
	the fabricated glass.
	The fitted spectrum corresponds to the parameters $\Gamma_0=1850~{\rm cm^{-1}}$, $\Gamma_1=240~{\rm cm^{-1}}$,
	$\Gamma_2=180~{\rm cm^{-1}}$, $\gamma_1=250~{\rm cm^{-1}}$, and $\gamma_2=190~{\rm cm^{-1}}$.
	Middle: Correlation diagram of energy levels of d$^3$ ions as a function of $Dq/(Dq+B)$. The rectangle indicates
	the Cr$^{3+}$-doped phosphate glass absorption bands.
	Right: Auxiliary Tanabe-Sugano diagram for ${\rm d}^3(O_h)$. The parameters $Dq$ and $B$ are obtained from
	Equations \eqref{eq:Dq} and \eqref{eq:B}. Then, the energy ratio $E(1{\rm ^2E})/E(1{\rm ^4T_2})$ at $Dq/(Dq+B)$
	determines the intersection with the $C/B$ curve. The intersection of the orange lines $Dq/(Dq+B)=0.683$
	(vertical) and $E(1{\rm ^2E})/E(1{\rm ^4T_2})=0.955$ (horizontal), indicated with the blue dot close to the
	curve $C/B=4.5$, accounts for the transition energies of the glass fabricated in this work, 
	$E({\rm ^2E}) = 14577~{\rm cm^{-1}}$, $E({\rm ^4T_2}) = 15256~{\rm cm^{-1}}$, and $E({\rm ^4T_1}) = 22148~{\rm cm^{-1}}$.}
	\label{fig1}
\end{figure*}

\begin{figure}[h]
\includegraphics[width=0.5\textwidth]{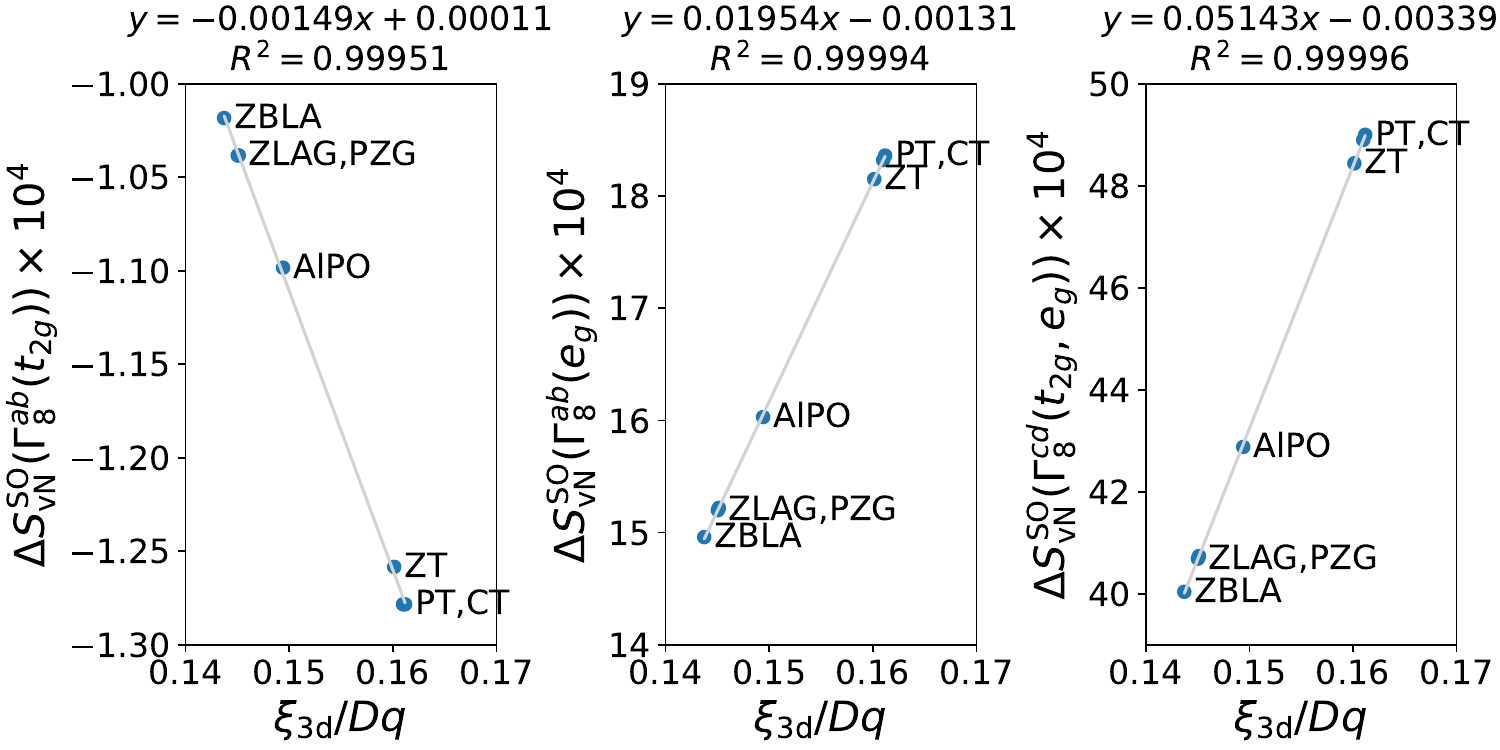}
	\caption{Spin-orbital entanglement entropy of $\Gamma_8$ espinors of Cr$^{3+}$-doped glasses versus $\xi_{\rm 3d}/Dq$
	ratio. ZLAG, ZBLA and PZG stand for fluoride glasses; PT, CT, and ZT stand for tellurite glasses;
	AlPO stands for aluminum-phosphate glass.}\label{lincorr}
\end{figure}

Figure \ref{fig1} shows the experimental and fitted optical absorption spectrum of the aluminum phosphate glass doped
with $\rm 1~mol~\%$ $\rm Cr^{3+}$. The energy correlation and Auxiliary Tanabe-Sugano diagrams for $\rm d^3$ ions are also
shown. These type of energy-level correlation diagrams are widely used in coordination chemistry for a fundamental understanding
of the interplay between electron-electron repulsion and crystal field strength \cite{SuganoBook,Garcia2023,Adachi2025}.
The absorption spectrum exhibits two main bands centered at $15256~{\rm cm^{-1}}$ ($\lambda=655.5~{\rm nm}$) and
$22148~{\rm cm^{-1}}$ ($\lambda = 451.5~{\rm nm}$), corresponding to the states ${\rm ^4T_2}$ and ${\rm ^4T_1}$, respectively.
The first band displays two dips, according to interferences with states ${\rm ^2E}$ and ${\rm ^2T_1}$, while the second band
exhibits a little shoulder which we assign to the state  ${\rm ^2T_2}$.
From these results and using Eqs. \eqref{eq:Dq} and \eqref{eq:B} we obtain the crystal field strength $Dq=1526~{\rm cm^{-1}}$
and the Racah parameter $B=723~{\rm cm^{-1}}$.
The ratio $Dq/(Dq+B)=0.683$ with $E(1{\rm ^2E})/E(1{\rm ^4T_2}) = (hc/686.0~{\rm nm})/(hc/655.5~{\rm nm}) = 0.955$
yields $4.0\le C/B \le 4.5$ (see Auxiliary TS diagram in Fig. \ref{fig1}). Then, varying $C$ from $4.0B = 2892~{\rm cm^{-1}}$
to $4.5B = 3254~{\rm cm^{-1}}$ with steps of $1~{\rm cm}^{-1}$ in the matrix \eqref{eq:Matrix}, the optimal value
$C=3108~{\rm cm^{-1}}$ is found which verifies $E(1{\rm ^2E})-E(1{\rm ^4A_2})=hc/686~{\rm nm}$.
The ratio $Dq/(Dq+B)=0.683$, marked with a rectangle in the correlation diagram (Fig. \ref{fig1}), confirms the participation
of the spin-forbidden states $\rm ^2E$, $\rm ^2T_1$, and $\rm ^2T_2$ in the absorption spectrum.
The fitting parameters, $\Gamma_1=240~{\rm cm^{-1}}$, $\Gamma_2=180~{\rm cm^{-1}}$, $\gamma_1=250~{\rm cm^{-1}}$, and
$\gamma_2=190~{\rm cm^{-1}}$, result in a one-electron spin-orbit coupling constant of $\xi_{\rm 3d}=228~{\rm cm^{-1}}$.
After calculating $\xi_{\rm 3d}$ and $Dq$, we proceed to calculate the amplitudes of the one-electron spinors
and the corresponding entanglement entropies. Using $\xi_{\rm 3d}=228~{\rm cm}^{-1}$, $Dq=1526~{\rm cm}^{-1}$, and $p/q=1$,
we obtain $\delta=0.242$ which yields $x_-=-y_+=-0.618$ and $y_-=x_+=0.786$. These amplitudes reveal a strong mixing of the
atomic spinors $3{\rm d}_{3/2}$ and $3{\rm d}_{5/2}$. This is because in $3{\rm d}^n$ systems, the spin-orbit interaction is
small compared to the strength of the crystal field; therefore, the crystal field spinors are mixed to represent the crystal
field spin-orbitals, which ultimately turn out to be a reasonable, albeit non-entangled, basis.
The entanglement entropies are presented in Table \ref{tab1}, which also includes the spin-orbit coupling constant
and the crystal field parameters for a family of chromium-doped glasses. The Neuhauser's parameters obtained from
the optical absorption spectra of fluoride \cite{Maalej2016} and tellurite glasses \cite{Taktak2021} doped with
Cr$^{3+}$ will be used for the theoretical study in this article.
The molar composition of fluoride glasses are \cite{Maalej2016}:
ZLAG(70.2 ZrF$_4$, 23.4 LaF$_3$, 0.6 AlF$_3$, 5.8 GaF$_3$, 0.5 CrF$_3$),
ZBL(57 ZrF$_4$, 5 LaF$_3$, 3.5 AlF$_3$, 34 BaF$_2$, 0.5 CrF$_3$), and
PZG(2 AlF$_3$, 34.3 GaF$_3$, 35.3 PbF$_2$, 4.9 YF$_3$, 23.5 ZnF$_2$, 0.5 CrF$_3$); while the tellurite glasses are made of
\cite{Kanth2005,Taktak2021} ZnF$_2$-MO-TeO$_2$ with MO=Pb labeled as PT, MO=CdO labeled as CT, and MO=ZnO labeled as ZT,
and the Cr$^{3+}$ content was varied from 0.1~\% to 0.3~\%.
A notable observation in our calculations is the behavior of the entanglement entropy for the $\Gamma_8^{ab}(t_{2g})$
Kramers pairs. To understand why $\Delta S_{\rm vN}^{\rm SO}$ (relative to a reference state or the non-relativistic limit)
can be negative or lower than expected, one must consider the symmetry constraints imposed by the octahedral field.
While the $\Gamma_7$ doublet arises from a specific combination of the $j=5/2$ manifold (see Eqs. \eqref{eq:G7a} and \eqref{eq:G7b}),
the $\Gamma_8$ quartet represents a more complex mixing of orbital and spin states. In the case of the $\Gamma_8^{ab}(t_{2g})$ pairs,
the crystal field acts to structure the orbital contribution. As the crystal field strength $Dq$ increases relative to the spin-orbit
coupling $\xi_{\rm 3d}$, the wavefunction is increasingly forced into a specific orbital symmetry that is more separable from the
spin than in the free-ion limit. Therefore, a negative $\Delta S_{\rm vN}^{\rm SO}$ does not imply a negative entropy in
an absolute sense, but rather a reduction in entanglement compared to the purely relativistic atomic limit. This reduction
occurs because the strong electrostatic environment effectively quenches a portion of the orbital angular momentum, partially
decoupling it from the spin and leading to a more localized, less entangled state.
In general, the entanglement entropy does not correlate with any of the Racah or crystal field strength parameters.
On the other hand, the spin-orbit coupling constant generally appears to correlate with the entanglement entropy,
except in the case of tellurite glasses, where the differences are small, if not negligible. In a family of $\rm 5d^1$
transition metal perovskites and hexachlorides, it has been shown that entanglement entropy correlates well with the effective
magnetic dipole moment \cite{GarciaRojas2025}.
\begin{table*}[t]
	\caption{Spin-orbit coupling constants, crystal field parameters, and spin-orbital entanglement entropies
	for a family of Cr$^{3+}$-doped glasses. ZLAG, ZBLA and PZG comprise fluoride glasses \cite{Maalej2016};
	PT, CT and ZT comprise tellurite glasses with composition ZnF$_2$-RO-TeO$_2$ (R = Pb,Cd and Zn) \cite{Taktak2021};
	AlPO refers to aluminum-doped phosphate glass.}
	\begin{tabular}{cccccccc}\hline\hline
		Glass system & $\xi_{\rm 3d}/{\rm cm^{-1}}$ & $Dq/{\rm cm^{-1}}$ & $B/{\rm cm^{-1}}$ & $C/{\rm cm^{-1}}$ &
										\multicolumn{3}{c}{$\Delta S_{\rm vN}^{\rm SO}/10^{-4}$} \\
			     &  &  &  &  & $\Gamma_{8}^{ab}(t_{2g})$ & $\Gamma_{8}^{ab}(e_{g})$ & $\Gamma_{8}^{cd}(t_{2g},e_g)$ \\ \hline
		ZLAG Ref. \cite{Maalej2016} & 209 & 1441 & 797 & 3156 &$-1.04$ & 15.20 & 40.69 \\
		ZBLA Ref. \cite{Maalej2016} & 213 & 1482 & 815 & 3084 &$-1.02$ & 14.96 & 40.04 \\
		PZG Ref. \cite{Maalej2016}  & 218 & 1502 & 762 & 3124 &$-1.04$ & 15.22 & 40.74 \\
		AlPO (this work) 	    & 228 & 1526 & 723 & 3108 &$-1.11$ & 16.03 & 42.88 \\
		PT Ref. \cite{Taktak2021}   & 240 & 1489 & 770 & 2719 &$-1.28$ & 18.36 & 49.00 \\
		CT Ref. \cite{Taktak2021}   & 241 & 1497 & 781 & 2748 &$-1.28$ & 18.32 & 48.90 \\
		ZT Ref. \cite{Taktak2021}   & 241 & 1505 & 790 & 2721 &$-1.26$ & 18.15 & 48.44 \\		
		\hline\hline
	\end{tabular}\label{tab1}
\end{table*}
The central finding of this work is the relationship between the electronic structure parameters and the degree of
spin-orbital entanglement. The entanglement entropy does not exhibit a simple or monotonic dependence on
$\xi_{\rm 3d}$, $Dq$, or the Racah parameters $B$ or $C$ independently. This suggests that the information-theoretic
complexity of the \ch{Cr^3+} electronic state is not dictated by the absolute magnitude of any single interaction.
Certainly, a correlation between the $\xi_{\rm 3d}/Dq$ ratio and the entanglement entropy mediated by the parameter
$\delta$ is expected, see Eq \eqref{eq:delta}.
Moreover, a robust linear correlation emerges when considering the dimensionless ratio $\xi_{\rm 3d}/Dq$.
Figure \ref{lincorr} shows the entanglement entropy $\Delta S_{\rm vN}^{\rm SO}(\Gamma_8^{cd}(t_{2g},e_g))$ versus $\xi_{\rm 3d}/Dq$
for the glasses analyzed here. As the ratio of the spin-orbit coupling to the crystal field strength increases,
the amount of entanglement entropy $\Delta S_{\rm vN}^{\rm SO}$ grows linearly. This linear behavior indicates that
the entanglement is a result of the competition between the local electrostatic environment, which tends to quench
the orbital angular momentum, and the relativistic effects that couple it back to the spin. In the regime where
$\xi_{\rm 3d}/Dq$ is small, the crystal field dominates, resulting in a state with low entanglement. As this ratio
increases, the mixing of the spin and orbital degrees of freedom becomes more pronounced, leading to the observed
linear increase in $\Delta S_{\rm vN}^{\rm SO}$.
This can be proof as follows: since we are dealing with a $\rm 3{\rm d}$ transition metal ion, the strong field
approximation $\xi_{3{\rm d}}/10Dq \ll 1$ is valid, suggesting the one-electron spinors can be recast in terms
of the standard crystal field spin-orbitals \cite{Stamokostas2018} $(t_{2g} \cup e_g) \otimes (\alpha \cup \beta)$, see Appendix. 
Indeed, when the crystal field spinors are rewritten in terms of the crystal field spin-orbitals, it is found that the contribution
of the crystal field orbital $e_g$ to the ground state, $(\sqrt{15}x_-+\sqrt{10}y_-)^2/25$, barely reaches 0.03 \%. Therefore, the
ground state multiplet ${\rm ^4A}_{2g}$ is well represented by the configuration $t_{2g}^{3}$, from which the spin-only formula for
the magnetic dipole moment is expected to be valid. This is consistent with the magnitude of the entanglement entropy found here,
which turns out to be two orders of magnitude smaller than those obtained for the $\rm 5d^1$ double perovskites Ba$_2$MgReO$_6$
and Ba$_2$NaOsO$_6$, whose entropies $\Delta S_{\rm vN}^{\rm SO}(\Gamma_8^{cd}(t_{2g},e_g))$ are 0.2112 and 0.2094,
respectively \cite{GarciaRojas2025}.
From Figure \ref{lincorr} we also observe that entanglement entropy serves as a descriptor of the nature of the host, i.e.,
it classifies glasses into families: tellurite glasses with the highest entropy, followed by phosphate glass, and then
fluoride glasses with the lowest entropy.

\section{\label{Conclusion}Conclusion}
In this work, we have presented a relativistic framework for determining the one-electron spinors of \ch{Cr^3+}
ions in glasses using optical absorption data. By applying this method to an aluminum phosphate glass system, we
demonstrated that the spin-orbital entanglement can be quantitatively assessed through the spin-orbital von Neumann
entropy, $\Delta S_{\rm vN}^{\rm SO}$. The developed framework is timely taking into account the recent interest
in detecting spin-orbital entanglemet \cite{Shen2026,Sun2026}. Our findings reveal that the entanglement entropy
does not depend on any single electronic parameter in isolation. Instead, a robust linear correlation was identified
with the dimensionless ratio $\xi_{\rm 3d}/Dq$. This suggests that the degree of spin-orbital mixing is governed by
the fundamental competition between relativistic effects and the local electrostatic environment. Specifically, for
$\Gamma_8^{ab}(t_{2g})$ states, the observed reduction in entropy ($\Delta S_{\rm vN}^{\rm SO}<0$) relative to
the atomic limit highlights the role of the crystal field in quenching orbital angular momentum and partially
restoring the separability of the spin and orbital degrees of freedom. The utility of the spin-orbital entanglement
entropy $\Delta S_{\rm vN}^{\rm SO}$ extends beyond a purely theoretical metric. It provides a singular, quantitative
measure of the relativicity of an electronic state within a specific chemical environment. By capturing how
interactions fundamentally alter the identity of the electronic degrees of freedom, $\Delta S_{\rm vN}^{\rm SO}$
serves as a diagnostic tool for identifying regimes where non-relativistic approximations fail. Furthermore,
since this entanglement directly influences magnetic anisotropy and zero-field splitting, the entropy framework
offers a new perspective for the rational design of transition metal-doped materials with tailored magneto-optical
properties. The ratio between the spin-orbital coupling constant and the strength of the crystal field $\xi_{\rm 3d}/Dq$
turns out to be a criterion for discerning the chemical nature of the host glass. It would be interesting to
extent the analysis of the spin-orbital entanglement not only to other types of glasses doped with Cr$^{3+}$,
but also to other materials, such as \ch{K_3Cr_xAl_{1-x}F_6} and \ch{K_3Cr_xGa_{1-x}F_6}, and \ch{Na3Al2Li3F12}:\ch{Cr^3+},
in which the the UV-Vis absorption spectrum also exhibits the interference pattern due to the spint-orbit interaction
\cite{Lee2020,Zhou2026}. Work along this line is currently in progress.

\begin{acknowledgements}
There is no funding to report in this work.
J. S. Robles-P\'aez and A. T. Carre\~no-Santos prepared the chromium-doped phosphate glass.
J. F. P\'erez-Torres conceived the original idea, developed the theoretical framework, and
performed the calculations. V. Garc\'ia-Rojas acquired the resources and was in charge of
the project. All the authors discussed the results and contributed to the final manuscript.
\end{acknowledgements}

\section*{Data Availabity Statement}
All the numerical resulst shown in this work can be obtained from the formulas and parameters described in the text.
The optical absorption spectrum of aluminum-phosphate doped-chromium glass is available from the corresponding author
upon reasonable request.

\appendix
\section{Basis transformations}\label{secAppendix}
\begin{widetext}
The crystal field orbitals mix in order to recover the atomic orbitals according to
\begin{equation}
	\begin{bmatrix}
		|n{\rm d}_{-2}\rangle \\
		|n{\rm d}_{-1}\rangle \\
		|n{\rm d}_{0}\rangle \\
		|n{\rm d}_{+1}\rangle \\
		|n{\rm d}_{+2}\rangle 
	\end{bmatrix}
	=
\begin{bmatrix}0 & \frac{\sqrt{2}}{2} & 0 & \frac{\sqrt{2}}{2} & 0\\1 & 0 & 0 & 0 & 0\\0 & 0 & 0 & 0 & 1\\0 & 0 & 1 & 0 & 0\\0 & - \frac{\sqrt{2}}{2} & 0 & \frac{\sqrt{2}}{2} & 0\end{bmatrix}	
	\begin{bmatrix}
		|t_{2g}^-\rangle \\
		|t_{2g}^0\rangle \\
		|t_{2g}^+\rangle \\
		|e_g^a\rangle \\
		|e_g^b\rangle 
	\end{bmatrix}
\end{equation}
Assuming $F_{n{\rm d}_{5/2}}(r) \approx F_{n{\rm d}_{3/2}}(r)$, where $F_{n{\rm d}_j}(r)$ is the radial part of the large component
of the atomic spinors \cite{GreinerBook}, the atomic states can be written in terms of the atomic spin-orbitals:
\begin{equation}
\begin{bmatrix}
	|n2{-\tfrac{3}{2}}\rangle \\
	|n2{-\tfrac{1}{2}}\rangle \\
	|n2{+\tfrac{1}{2}}\rangle \\
	|n2{+\tfrac{3}{2}}\rangle \\
	|n{-3}{+\tfrac{5}{2}}\rangle \\
	|n{-3}{-\tfrac{5}{2}}\rangle \\
	|n{-3}{-\tfrac{3}{2}}\rangle \\
	|n{-3}{-\tfrac{1}{2}}\rangle \\
	|n{-3}{+\tfrac{1}{2}}\rangle \\
	|n{-3}{+\tfrac{3}{2}}\rangle 
\end{bmatrix}
=
\begin{bmatrix}- \frac{2 \sqrt{5}}{5} & 0 & 0 & 0 & 0 & 0 & \frac{\sqrt{5}}{5} & 0 & 0 & 0\\0 & - \frac{\sqrt{15}}{5} & 0 & 0 & 0 & 0 & 0 & \frac{\sqrt{10}}{5} & 0 & 0\\0 & 0 & - \frac{\sqrt{10}}{5} & 0 & 0 & 0 & 0 & 0 & \frac{\sqrt{15}}{5} & 0\\0 & 0 & 0 & - \frac{\sqrt{5}}{5} & 0 & 0 & 0 & 0 & 0 & \frac{2 \sqrt{5}}{5}\\0 & 0 & 0 & 0 & 1 & 0 & 0 & 0 & 0 & 0\\0 & 0 & 0 & 0 & 0 & 1 & 0 & 0 & 0 & 0\\\frac{\sqrt{5}}{5} & 0 & 0 & 0 & 0 & 0 & \frac{2 \sqrt{5}}{5} & 0 & 0 & 0\\0 & \frac{\sqrt{10}}{5} & 0 & 0 & 0 & 0 & 0 & \frac{\sqrt{15}}{5} & 0 & 0\\0 & 0 & \frac{\sqrt{15}}{5} & 0 & 0 & 0 & 0 & 0 & \frac{\sqrt{10}}{5} & 0\\0 & 0 & 0 & \frac{2 \sqrt{5}}{5} & 0 & 0 & 0 & 0 & 0 & \frac{\sqrt{5}}{5}\end{bmatrix}
\begin{bmatrix}
	|n{\rm d}_{-2}\alpha\rangle \\
	|n{\rm d}_{-1}\alpha\rangle \\
	|n{\rm d}_{0}\alpha\rangle \\
	|n{\rm d}_{+1}\alpha\rangle \\
	|n{\rm d}_{+2}\alpha\rangle \\
	|n{\rm d}_{-2}\beta\rangle \\
	|n{\rm d}_{-1}\beta\rangle \\
	|n{\rm d}_{0}\beta\rangle \\
	|n{\rm d}_{+1}\beta\rangle \\
	|n{\rm d}_{+2}\beta\rangle 
\end{bmatrix}
\end{equation}
Then, the crystal field spinors,
\begin{equation}
	\begin{bmatrix}
		|\Gamma_{8\pm}^a\rangle \\
		|\Gamma_{8\pm}^b\rangle \\
		|\Gamma_{8\pm}^c\rangle \\
		|\Gamma_{8\pm}^d\rangle \\
		|\Gamma_7^a\rangle \\
		|\Gamma_7^b\rangle 
	\end{bmatrix}
	=
\begin{bmatrix}0 & 0 & 0 & y & 0 & \frac{\sqrt{30} x_\pm}{6} & 0 & 0 & 0 & \frac{\sqrt{6} x_\pm}{6}\\- y_\pm & 0 & 0 & 0 & \frac{\sqrt{30} x_\pm}{6} & 0 & \frac{\sqrt{6} x_\pm}{6} & 0 & 0 & 0\\0 & y_\pm & 0 & 0 & 0 & 0 & 0 & x_\pm & 0 & 0\\0 & 0 & - y_\pm & 0 & 0 & 0 & 0 & 0 & x_\pm & 0\\0 & 0 & 0 & 0 & 0 & \frac{\sqrt{6}}{6} & 0 & 0 & 0 & - \frac{\sqrt{30}}{6}\\0 & 0 & 0 & 0 & \frac{\sqrt{6}}{6} & 0 & - \frac{\sqrt{30}}{6} & 0 & 0 & 0\end{bmatrix}
\begin{bmatrix}
	|n2{-\tfrac{3}{2}}\rangle \\
	|n2{-\tfrac{1}{2}}\rangle \\
	|n2{+\tfrac{1}{2}}\rangle \\
	|n2{+\tfrac{3}{2}}\rangle \\
	|n{-3}{+\tfrac{5}{2}}\rangle \\
	|n{-3}{-\tfrac{5}{2}}\rangle \\
	|n{-3}{-\tfrac{3}{2}}\rangle \\
	|n{-3}{-\tfrac{1}{2}}\rangle \\
	|n{-3}{+\tfrac{1}{2}}\rangle \\
	|n{-3}{+\tfrac{3}{2}}\rangle 
\end{bmatrix}
\end{equation}
can be written in terms of the crystal field spin-orbitals:
\begin{eqnarray}
	|\Gamma_{8\pm}^a\rangle &=&
		\left(\frac{\sqrt{30}x_\pm}{15}-\frac{\sqrt{5}y_\pm}{5}\right)|t_{2g}^+\alpha\rangle
		+\left(\frac{2\sqrt{15}x_\pm}{15}-\frac{\sqrt{10}y_\pm}{5}\right)|t_{2g}^0\beta\rangle 
		+\left(\frac{\sqrt{15}x_\pm}{5}+\frac{\sqrt{10}y_\pm}{5}\right)|e_{g}^a\beta\rangle \\
	|\Gamma_{8\pm}^b\rangle &=&
		\left(\frac{\sqrt{30}x_\pm}{15}-\frac{\sqrt{5}y_\pm}{5}\right)|t_{2g}^-\beta\rangle
		-\left(\frac{2\sqrt{15}x_\pm}{15}-\frac{\sqrt{10}y_\pm}{5}\right)|t_{2g}^0\alpha\rangle
		+\left(\frac{\sqrt{15}x_\pm}{5}+\frac{\sqrt{10}y_\pm}{5}\right)|e_{g}^a\alpha\rangle \\
	|\Gamma_{8\pm}^c\rangle &=&
		\left(\frac{\sqrt{10}x_\pm}{5}-\frac{\sqrt{15}y_\pm}{5}\right)|t_{2g}^-\alpha\rangle
		+\left(\frac{\sqrt{15}x_\pm}{5}+\frac{\sqrt{10}y_\pm}{5}\right)|e_{g}^b\beta\rangle \\
	|\Gamma_{8\pm}^d\rangle &=&
		\left(\frac{\sqrt{10}x_\pm}{5}-\frac{\sqrt{15}y_\pm}{5}\right)|t_{2g}^+\beta\rangle
		+\left(\frac{\sqrt{15}x_\pm}{5}+\frac{\sqrt{10}y_\pm}{5}\right)|e_{g}^b\alpha\rangle \\
	|\Gamma_{7}^a\rangle &=& -\frac{\sqrt{6}}{3}|t_{2g}^+\alpha\rangle + \frac{\sqrt{3}}{3}|t_{2g}^0\beta\rangle \\
	|\Gamma_{7}^b\rangle &=& - \frac{\sqrt{6}}{3}|t_{2g}^-\beta\rangle -\frac{\sqrt{3}}{3}|t_{2g}^0\alpha\rangle
\end{eqnarray}
\end{widetext}

\nocite{*}
\bibliography{bibliography}

\end{document}